# Dynamics of Nonequilibrium Dimerization of Model Polymer Chains

**Sangita Mondal,[†] Ved Mahajan[#] and Biman Bagchi [†] ***

[†] *SSCU, Indian Institute of Science, Bangalore 560012, India.*
[#] *Department of Chemistry, Indian Institute of Science Education and Research (IISER) Tirupati, Tirupati, Andhra Pradesh 517619, India*
*Authors to whom correspondence should be addressed: bbagchi@iisc.ac.in

## Abstract

**Dimerization and subsequent aggregation of polymers and biopolymers often occur under nonequilibrium conditions. When the initial state of the polymer is not collapsed, or the final folded native state, the dynamics of dimerization can follow a course sensitive to both the initial conditions and the conformational dynamics. Here we study dimerization process by using computer simulations and analytical theory where both the two monomeric polymer chains are in the elongated state and are initially placed at separation distance, $d_0$. Subsequent dynamics lead to the concurrent processes of collapse, dimerization and/or escape. We employ Langevin dynamics simulations with a coarse-grained model of the polymer to capture certain aspects of the dimerization process. At separations $d_0$ much shorter than the length of the monomeric polymer, the dimerization could happen fast and irreversibly, from the partly extended polymer state itself.** *At an initial separation larger than a critical distance, $d_c$, the polymer collapse precedes dimerization and a significant number of single polymers do not dimerize within the time scale of simulations.* **To quantify these competition, we introduce several time-dependent order parameters, namely, (i) the time-dependent radius of gyration $R_G(t)$ of individual polymers describing the conformational state of the polymer, (ii) a centre-to-centre of mass distance parameter $R_{MM}$, and (iii) a time dependent overlap function Q(t) between the two monomeric polymers, mimicking contact order parameter popular in protein folding. In order to better quantify the findings, we perform a theoretical analysis to capture the stochastic processes of collapse and dimerization by using dynamical disorder model.**

# I. Introduction

Dimerization of polymers in solution is a complex process controlled by many factors, including conformational fluctuations of the individual monomers, their mutual diffusion coefficient, concentration in solution, and the nature of the solvent.[1–3] The equilibrium conformation of a single polymer chain is determined by the nature of the polymer-solvent interaction. In a good solvent, the polymer remains in a swollen state, while in a poor solvent, it collapses.[4,5] The final configuration of the polymer, combined with the concentration that determines the relative separation, could play an important role in polymer dimerization. In the real world, polymers can dimerize from a non-equilibrium state. In that situation, conformational fluctuations could play an even larger role when the monomers are in the equilibrium state. It is actually an interesting problem to investigate the interplay between the factors that determine polymer dimerization.

Macromolecules self-aggregate, form self-assembly, and can also undergo phase separation. These multitude of phenomena occur widely in nature under diverse conditions influenced by non-covalent interactions.[4,6–8] In recent years, scientists have increasingly focused on understanding these processes in biological polymers such as proteins.[9–12] Protein association and dissociation stand as important biochemical phenomena.[13,14] For instance, aggregation of amyloid β-fibrils poses a severe problem in the human body. It is implicated in the deadly Alzheimer's disease.[15] Similarly, an association of alpha-synuclein is implicated in Perkinson's disease.[14,16,17] Another important example is the formation of insulin hexamers and dimers, which play a crucial role in storing and delivering the hormone. Dimer formation and dissociation are intermediate steps in this process.[13]

Several theoretical, experimental, and computational investigations have explored the stability, kinetics, and mechanisms involved in the association and dissociation processes of protein oligomers.[18,19] Macromolecular self-assemblies also hold promise for applications such as surface engineering, tissue repair scaffolding, and the fabrication of conducting nanowires.[20,21].

In a dilute solution of macromolecules, dimerization is the early and important stage of the aggregation process. So, by using simple coarse-grained polymer models, one can mimic the macromolecular system and can generalize the results to a wide category of biological and synthetic polymers.[1,5] A large number of studies have extensively employed stick-bead or spring-bead model polymers as invaluable tools to understand the intricate behaviours of macromolecular systems. [22–25] Toy models have been used for a long time to understand polymer dynamics.

In the biological world and the polymer industry, we often start with monomeric polymers, which are synthesized by various processes. [26,27] These monomeric polymers vary widely in size, from a few tens of constituent repeat units (like residues in proteins) to many millions, as in the polymer industry. Often, these monomeric polymers form an aggregate in a process that is far more complex and intriguing than the aggregation of simple small molecules, like inorganic clusters formed by the condensation process. In the case of rather large monomeric polymers, fluctuations in shape and size and the availability and exposure of associative surfaces play important roles, in addition to the distance between the monomers. The collapse and dimerization of polymers refer to their ability to undergo conformational changes in response to changes in their environment or external stimuli. For example, Chakrabarty et al. illustrate through detailed atomistic simulations that the stable conformation of linear hydrocarbon (n-alkane) chains changes from an elongated state to a collapsed state with variations in chain length.[28] Also, the aggregation behavior of polymeric macromolecules depends on the flexibility of the polymer, and it is found that the dimerization probability is a non-monotonic function of the polymer persistence length.[29] In recent decades, efforts have been made to understand the structural characteristics and dynamic behavior of linear homopolymers during the process of collapse.[30–34] These systems confront us with an exceedingly large number of degrees of freedom to be tractable within analytically solvable models. As a result, computer simulation has proven to be a very powerful and useful tool in this field. A significant body of theoretical research has been carried out utilizing both implicit and explicit solvent models. In the realm of implicit solvent models, analytical, theoretical investigations span from extensions of the Flory-Huggins mean-field[4,35] approach to advanced field theoretic techniques. The quality or nature of the solvent enters implicitly through interaction between atoms of the polymer chain.

But, to the best of our knowledge, there is a lack of quantitative research on the role of polymer collapse on the dimerization dynamics of the bead-spring model, which is the central focus of our study. Our primary goal is to achieve a comprehensive understanding of the dynamic behaviours associated with polymer collapse and dimerization of two specified polymers.

While the association of collapsed and globular polymers is a subject of scientific and practical interest, few studies have been carried out on dimerization from an initial nonequilibrium state. One could imagine that if polymers (or proteins) are synthesized at a fast rate, then there exists a certain possibility that two polymers can associate even before they attain their final collapsed state. The rate of association in the extended state can be quite different from that in the collapsed state. This additional factor has not been thoroughly investigated in the published literature. An example of association from the extended state can be found in the droplet formation by association of intrinsically disordered proteins, a subject of intense interest at present.[36–38]

In the present study, we investigate the polymerization of two model polymers, which are initially in extended states, separated by a certain initial distance $d_0$. We employ Langevin dynamic simulation techniques to understand the process of non-equilibrium dimerization. Dimerization and subsequent aggregation of polymers and biopolymers often occur under nonequilibrium conditions. Configuration collapse of the individual chains is found to compete with the mutual diffusion and relative orientation of the two chains. The primary focus is to study the simultaneous occurrence of polymer dimerization and collapse. Even such a simple system offers interesting features. The polymeric monomers can associate or diffuse away, with almost no chance of returning once they are separated beyond a certain distance. The simulations show rich behavior and manifest correlations between collapse and dimerization. Here, we explore the specific limited parameter space, where the final equilibrium state of the monomeric single polymer chain is the collapsed state. Thus, we consider the poor solvent limit where the intermolecular interaction facilitates the polymer collapse. We developed a stochastic theory that can elucidate and capture some aspects of the dimerization process. It can be seen that the theory needs to be far more sophisticated than earlier treatments that treated dimerization as the diffusion-controlled aggregation of spheres.

The paper is organized as follows. First, we will discuss the order parameters in section II. This is followed in section III, by describing the system and the simulation details. In section IV, we discuss the sequence of events as obtained from simulation. Section V contains tie dependence of order parameters as obtained from averaging over trajectories. In sections VI and VII, we discuss the critical distance calculation and time scale of the process. Further, we discuss the Dynamical disorder model in Section VIII, and then In Sec. IX, we end with the importance of these findings and also with some general conclusions.

## II. Order parameter-based description for polymer dimerization

A quantitative description of the dimerization of polymers requires the introduction of certain collective variables that can depict the progression of dimerization and separation of the two polymers involved. The use of such collective variables is quite common in phase transitions, as in Landau's theory [5,39,40]. Order parameters are known to play an important role not just in a theoretical description but also in providing valuable physical insight into a complex process. Order parameters allow us to make a transition from microscopic to physically realizable quantities where we can apply thermodynamic and kinetic rules.

To study the microscopic aspects and mechanism of polymer dimerization, we observed that the following four collective variables, detailed below, could provide the required quantification of the dimerization process.

(i) The distance between the center-of-mass of two monomeric units ($R_{MM}$).

(ii) The second order parameter is selected to describe the other competing process, namely, self-organization is quantified by using the radius of gyration, $R_G(t)$, which is defined below in Eq. (1)

$$R_G(t) = \sqrt{\frac{(x(t) - R_{com}(t))^2 + (y(t) - R_{com}(t))^2 + (z(t) - R_{com}(t))^2}{N}} \tag{1}$$

In Eq.(1), the *x, y,* and *z* are the three coordinates of the corresponding polymer. The reorganization happens when the polymer collapses from an extended state to a globular state. During this process, the radius of gyration of the polymer changes drastically. Hence, we use the radius of gyration as the order parameter to quantify the reorganization of the polymer.

(iii) A contact order parameter $Q(t)$ is an analog of the contact order parameter in protein folding. This quantifies the number of stable monomer-monomer pair contacts. Here, we need to define a contact as a topological event where one atom of a polymeric monomer is surrounded by at least two atoms of the other polymeric monomer.

(iv) In order to quantify the effective distance between the two polymeric monomers, we introduce another order parameter, $R_{CP}(t)$, defined as follows

$$R_{CP}(t) = \sqrt{\sum_{i,j}\left((x_{1i}-x_{2j})^2 + (y_{1i}-y_{2j})^2 + (z_{1i}-z_{2j})^2\right)} \quad (2)$$

In Eq. (2), $x_{1i}$, $y_{1i}$ and $z_{1i}$ are the x, y, and z coordinates of polymer *1*, respectively. Similarly, $x_{2i}$, $y_{2i}$ and $z_{2i}$ are the x, y, and z coordinates of polymer 2, respectively. The summation runs over all the monomers of polymer 1 and polymer 2. $R_{CP}(t)$ measures the collective distance between each monomer of polymer 1 from each monomer of polymer 2. The square root ensures that $R_{CP}(t)$ is of dimension of length. As the polymers form a dimer, the collective distance order parameter $R_{CP}(t)$ should exhibit a sharp decrease in its value. However, there could be transient contacts where the monomers are connected only at the ends but not anywhere else. This order parameter helps to pick up such structures.

The four-order parameters reflect the complexity of the polymer dimerization process. Note that as we address homopolymer dimerization, the description differs from heteropolymer dimerization as in proteins where pair contacts to be formed towards the final native state are defined a priori. In contrast, here the contact order parameter is not unique. and different dimerization events can proceed via different atomic contact formations between the monomers.

### III. Simulation details

We use a simple coarse-grained model of polymer study dimerization and reorganization. The model is described in **Figure 1.** We consider a flexible bead-spring polymer chain where the monomers are connected in a linear way. All the monomers interact via a Lennard-Jones potential,

$$V_{ij}^{LJ}(r_{ij}) = \begin{cases} 4\varepsilon\left[\left(\dfrac{\sigma}{r_{ij}}\right)^{12} - \left(\dfrac{\sigma}{r_{ij}}\right)^{6}\right], & r_{ij} \leq 2.5\sigma \\ 0, & r_{ij} > 2.5\sigma \end{cases} \quad (3)$$

Where $r_{ij}$ is the distance between the *i*-th and the *j*-th monomer. The Lennard-Jones parameters $\sigma$ and $\varepsilon$ were both set to 1. The monomers, which are the nearest neighbors along the chain, are connected by a non-linear spring with the potential.[41]

$$V_{FENE}(r_{ij}) = -\dfrac{1}{2}kR_0^2 \ln\left[1 - \left(\dfrac{r_{ij}}{R_0}\right)^2\right] \quad (4)$$

$V_{FENE}(r_{ij})$ is a potential that is characterized by two parameters, k, and $R_0$. k is the spring constant. $R_0$ corresponds to the maximum length up to which the bond can extend. k can extend till $R_0$.

This form of potential has been used several times to model bonded interaction.[42–46] We set the parameters $R_0 = 1.5\sigma$ and $k = \dfrac{30\varepsilon}{\sigma^2}$. The value of the spring constant was large enough to restrict the bond length such that it was always below $1.2\sigma$.

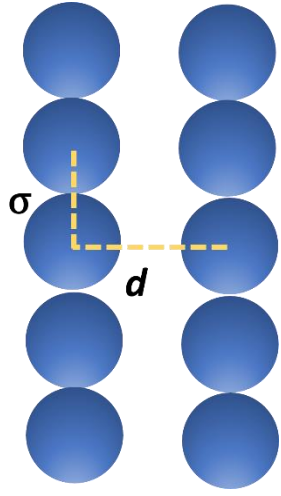

**Figure 1. Schematic diagram of bead spring model. The polymer consists of N number of beads, each representing a monomer. Each bead has a diameter σ and the separation between the two polymers is d. This distance needs to change to achieve mutual contact and dimerization.**

We use LAMMPS software to study polymer dimerization with time step $0.001\tau$.[47]

In our polymer system, the equation of motion of the particles is governed by the Langevin equation,[48,49]

$$m_i \frac{d^2 r_i}{dt^2} = -m_i \zeta_j \frac{dr_i}{dt} + F_i(t) + R_i(t) \tag{5}$$

where $m_i$ denotes the mass of particle $i$, $F_i$ is the interaction force acting on i-th particle by its environment, $\zeta_j$ is the friction coefficient, and $R_j(t)$ signifies the fluctuating or random force with $\langle R_i(t) \rangle = 0, \langle R_i(t) R_j(t') \rangle = 2\zeta \delta_{ij} k_B T \delta(t-t')$. Here, we have used reduced units. All parameters used in the simulation are given in tabular form.

**Table 1: Table of Parameter Sets and the relation between real unit and reduced unit.**

| | Reduced Unit | Real unit | Relation between real unit and reduced unit |
|---|---|---|---|
| Temperature | 1.0 | 119.8K | $T^* = \dfrac{k_B T}{\varepsilon}$ |
| Pressure | 1.0 | 41.9Mpa | $P^* = \dfrac{\sigma^3}{\varepsilon} P$ |
| Time | 1.0 | 2.15ps | $\tau^* = \tau \sqrt{\dfrac{\varepsilon}{m\sigma^2}}$ |

Here m the mass ($39.9 gm/mol$), σ the unit of length ($3.41 \text{Å}$), and ϵ the unit of energy $0.238 Kcal/mol$. Here, we explore a specific parameter space, where the equilibrium state of a single monomeric polymer chain is the collapsed state, driven by favourable intermolecular interactions. The next few sections present the results of our numerical simulations.

## IV. Sequence of events: Trajectory analysis

In Figures 2, Figure 3, Figure 4, and Figure 5, we show snapshots of the progression of events for d = 4σ, 8σ, 12σ and 16σ. The temporal evolution differs from small to large separation distance. We observe that at a small separation distance, i.e., d = 4σ, the end-to-end contact forms early, before the collapse, as depicted in **Figure 2(b).** Following this initial contact, the polymers proceed to dimerize, where they adopt an entangled helix-like structure, as shown in **Figure 2 (c-d)**. Subsequently, this entangled structure undergoes collapse, resulting in the formation of a globular structure, as shown in **Figure 2 (f).**

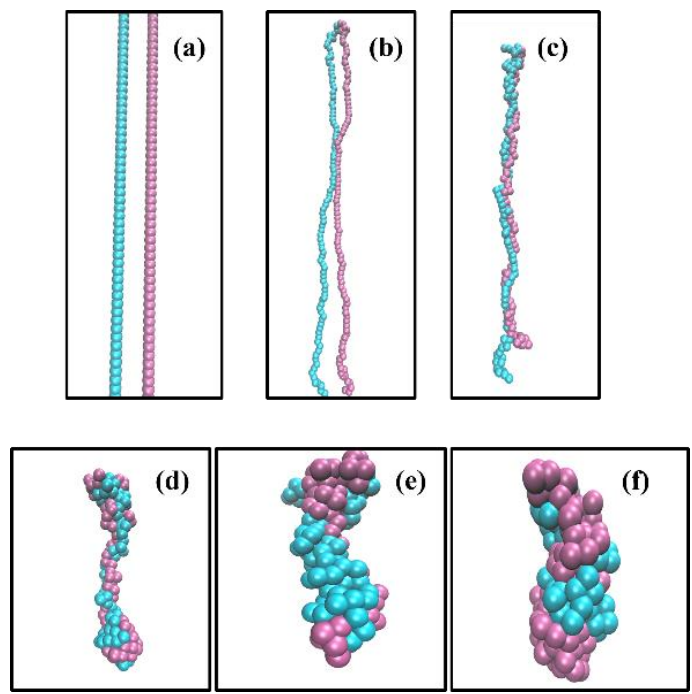

**Figure 2. Snapshot of the simulation trajectories polymer with initial separation distance d = 4σ. As the simulation progresses, the polymer starts to collapse (a)-(c) and form an entangled structure (d)-(e). End up with the merged compact collapsed state(g).**

Dimerization starting from separations 8σ and 12σ exhibits markedly different behaviour and presents interesting contrasts. At these two intermediate separations (d = 8σ and 12σ), we see the collapse of individual polymers and the formation of 'pearls', **Figure 3(b)** and **4 (b).** However, at separation of 8σ, the contacts are made partially, and collapse follows with the elongated dimeric state. On the other hand, at d = 12σ, the collapse takes place before dimerization. Here, the collapsed state undergoes fluctuations to make contact with the second polymer, which initiates the dimerization process. Subsequently, this entangled structure undergoes mixing, resulting in the formation of a globular structure, which is like a binary mixture.

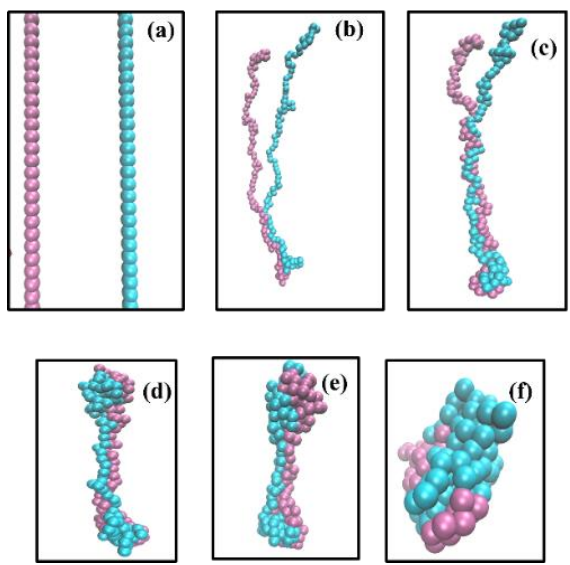

**Figure 3.** Snapshot of the simulation trajectories polymer with initial separation distance d = 8σ. As the simulation progresses, the polymer starts to collapse and form pearls (a)-(c). Further, it forms a compact structure.

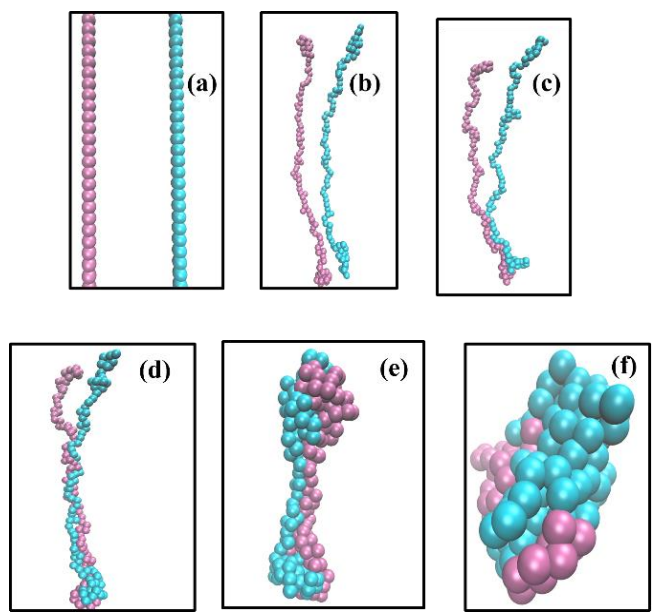

**Figure 4.** Snapshot of the simulation trajectories polymer d = 12σ. As the simulation progresses, the polymer starts to collapse (a)-(c) and form an entangled structure (d)-(e). End up with a compact collapsed state(g).

When the separation distance is large d =16σ, we observe that the two polymers collapse individually, as depicted in **Figure 5 (b-d)**, resembling the behavior of a single chain. There is no contact formation during this period. Subsequently, these collapsed polymers approach each other and form dimerized compact structures.

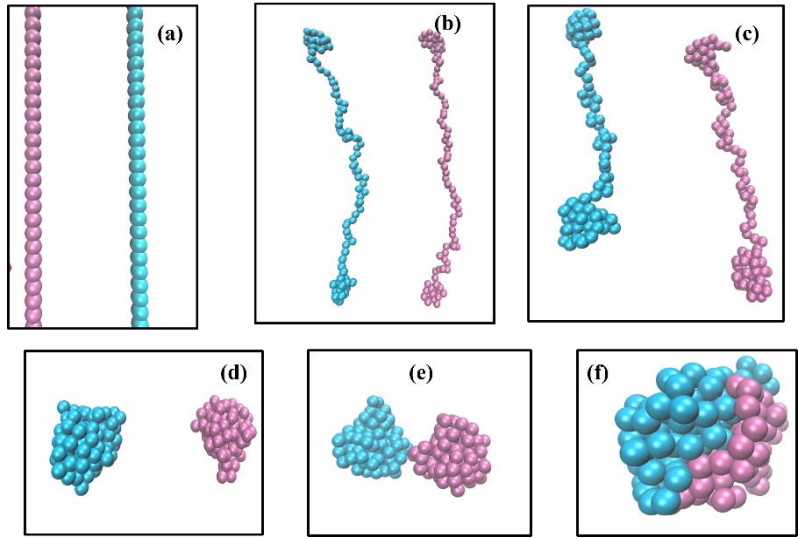

**Figure 5.** Snapshot of the simulation trajectories polymer with N = 100 and initial separation distance d = 16σ. The two polymers collapse individually, similar to a single chain, without any signature of dimerization. Further, the collapsed polymers approach each other and form dimerized structures in their collapsed states.

# V. Time dependence of order parameters

In this section, we have shown the time evolution of the order parameters defined above. The theoretical aspects of this problem acquire special interest because of the interaction between these two evolving order parameters that control the dimerization process. The entire process, although stochastic, exhibits predictable trends in the separation d and the length of the polymers. The other characteristics of the polymer, even the medium, can play important roles that are beyond the scope of this study. However, this combined simulation-theory study reveals a lot of rich dynamics.

**V.1. Time evolution of $R_G(t)$**

**Figure 6** shows the time evolution of $R_G(t)$ for four different initial separation distances. We find that the internal reorganization of the polymer, which happens through the collapse of the polymer from an extended state to a globular state, does not exhibit any dependence on the distance of separation d. This suggests that the dimerization does not affect the internal collapse of the polymer.

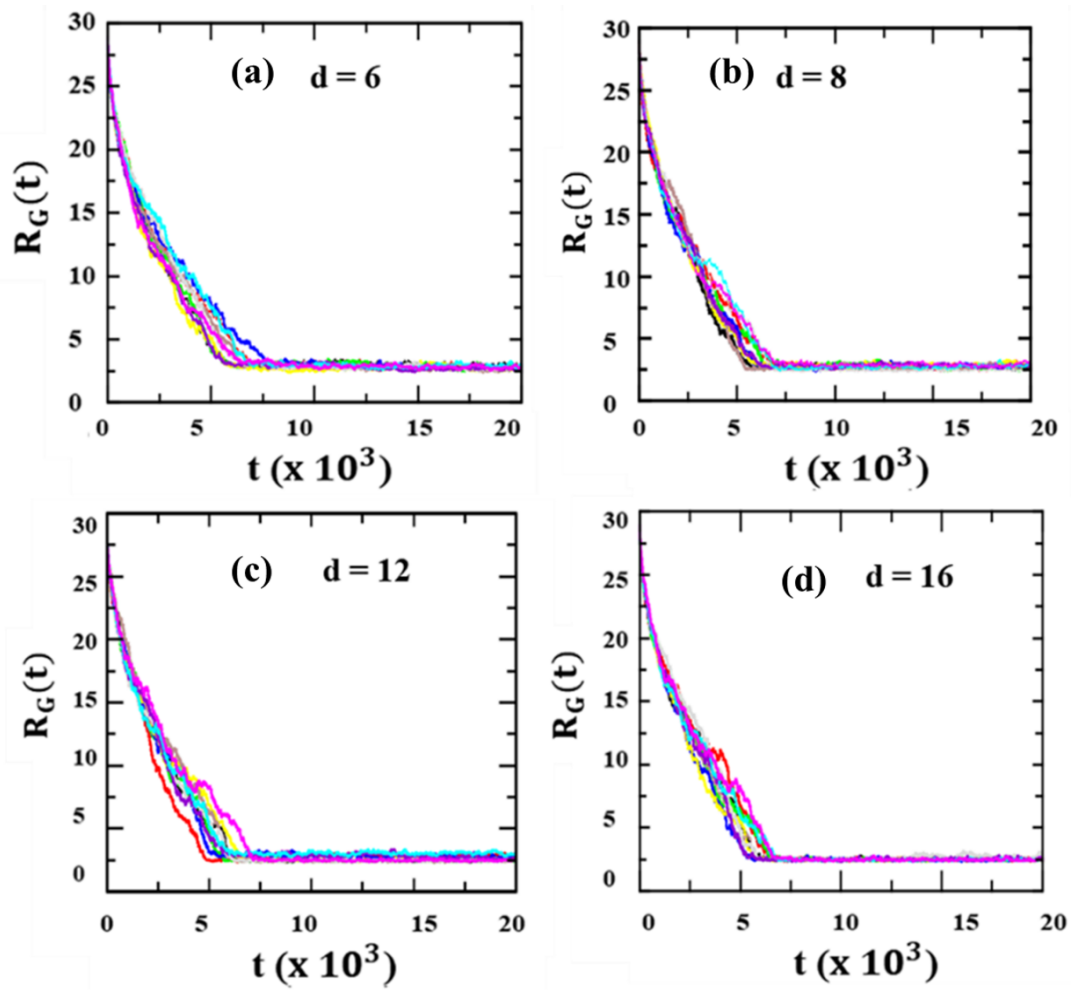

**Figure 6.** Time evolution of $R_G(t)$ for four different distances of separation d. We observe there is no such dependence of initial separation (d) on the radius of gyration.

### V.2. Time evolution of $R_{CP}(t)$

**Figure 7.** quantifies the dependence of the progression of polymer dimerization on the distance of initial separation $d$ between the two polymers. The progression of overlap is measured by the value of $R_{cp}(t)$. This figure shows that there is a threshold value of d ($d = 8\sigma$) below which the polymers almost always form a stable dimer and above which they remain separate. This suggests that the dimerization process is sensitive to the spatial relative arrangement of the polymers and can be controlled by changing the concentration. The figure also reveals that the dimerization process is fast and irreversible, as the value of R_cp(t) reaches a steady state within a short time and does not change when d is changed.

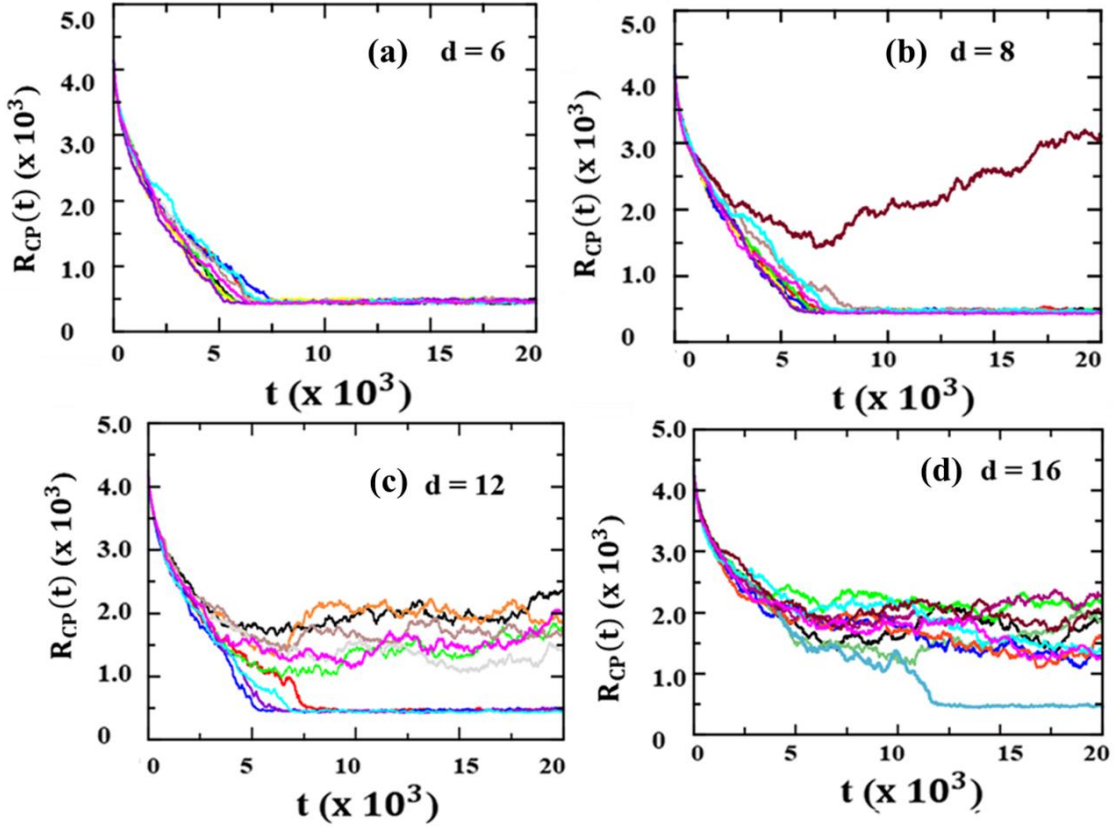

**Figure 7.** The time evolution $R_{cp}(t)$ for four increasing distances of separation. We see that when the polymers are separated up to a d = $8\sigma$, the value of $R_{CP}(t)$ attains a much lower stable value. This low value represents the dimerized state of the two polymers. However, if the distance between the two polymers is increased beyond $8\sigma$, $R_{CP}(t)$ values are significantly larger. This denotes the presence of a critical distance of separation beyond which dimerization does not occur.

Furthermore, in **Figure 8**, we show the trajectory of the system on the two-dimensional order parameter plane spanned by the two order parameters $R_G(t)$ and $R_{MM}(t)$. At long times, the individual polymers indeed undergo their expected collapse, but the distance of separation is larger than the critical separation dc, it does not show any association, as shown by the green curve. Whereas the separation distance is less than the critical separation dc, the polymers undergo dimerization (blue curve).

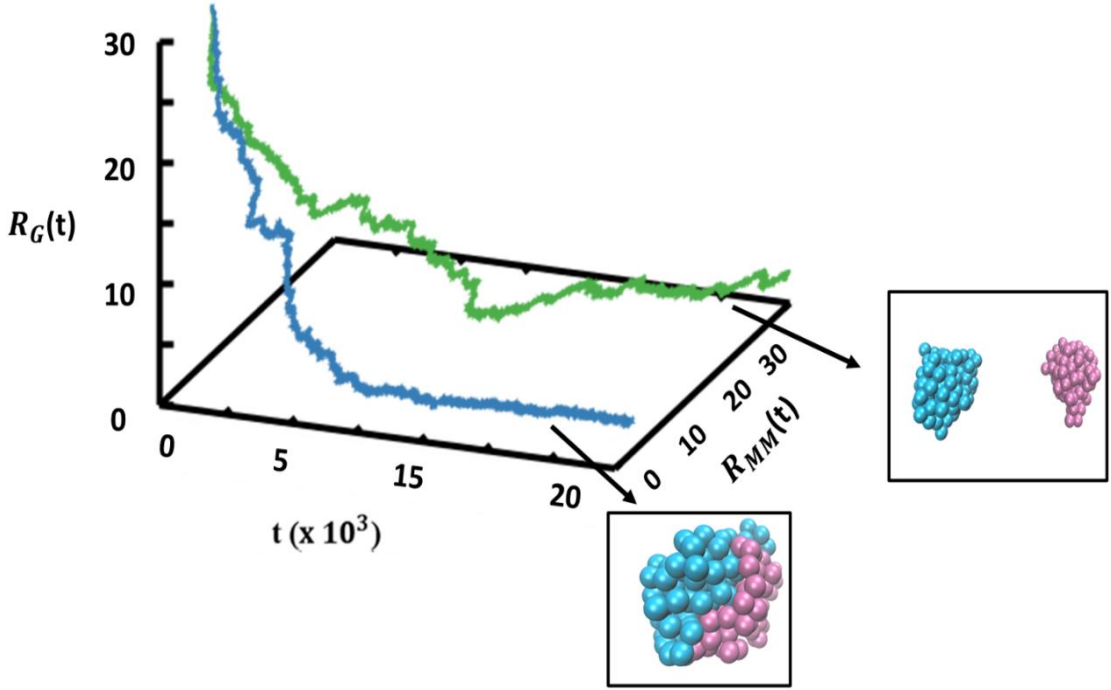

**Figure 8. The trajectory of the system on the two-dimensional order parameter space spanned by the radius of gyration $R_G(t)$ and $R_{MM}(t)$ is shown here for initial separation distance 8σ. When the separation distance exceeds the critical separation dc, there is no evident association, as depicted by the green curve. Conversely, when the separation distance is below dc, the polymers undergo dimerization (illustrated by the blue curve). Corresponding simulation snapshots are shown in the inset.**

## VI. Determination of the critical distance of separation

It is clear from the results we have presented so far that single-chain polymers undergo dimerization and collapse below a certain distance.

Further, to find this critical distance, we have calculated the time evolution of the center of mass distance between the two polymers with four different initial separation distances 6σ, 8σ, 12σ and 16σ. For each distance of separation, we perform ten independent simulations. **Figure 9** shows the distribution of the center of mass distance P($R_{MM}$) between the two polymers. For a separation distance of d = 6σ, a prominent peak is observed at low $R_{MM}$, indicating the presence of fully dimerized states. As the initial separation distance increases, the distribution becomes broader, and the height of the first peak decreases. Specifically, for d = 16σ, the center of mass distribution [P($R_{MM}$)] exhibits a broad profile at larger $R_{MM}$ values.

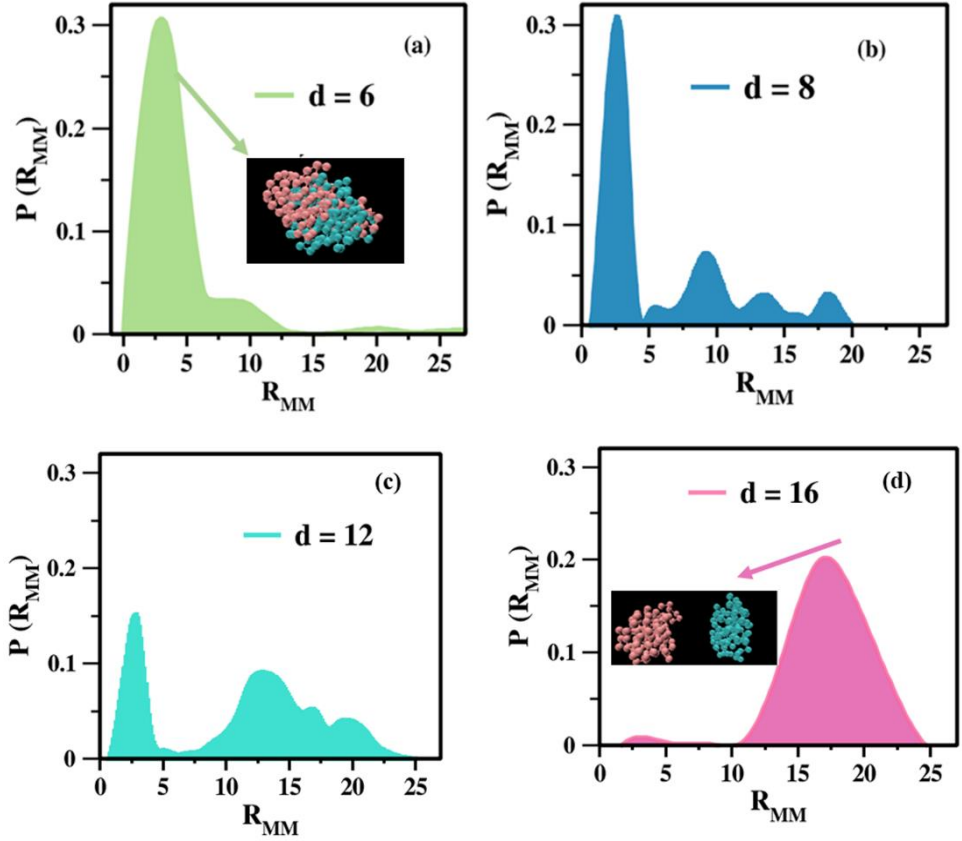

**Figure 9.** The distribution of the center of mass distance of two polymers at (a) d = 6σ, (b) d = 8σ, (c) = 12σ, and (d) d = 16σ. At a higher d, there is a broad peak at $R_{MM}$ = 15; however, as the initial separation distance is lowered, the peak shifts from large $R_{MM}$ to small $R_{MM}$, as shown in panels (a-c).

## VII. Calculation of reaction time distribution

Further, to get an idea of the time scale of the dimerization, we calculated the reaction time distribution function. At first, we compute the $R_{MM}(t)$, which measures the distance between the center of mass of the two polymers, $Q$ which measures the number of contacts. Further, we introduce two time constants $\tau_Q$ and $\tau_{MM}$, where $\tau_Q$ represents the time required for the two polymers to form at least $N/2$ number of contacts and $\tau_{MM}$ signifies the time taken by the two polymer to reach the centre of mass distance $2.5\sigma$. **Figure 10** shows the distribution of $\tau_Q$. We observe that as the separation increases (from d = 4 σ to 12σ), the peak progressively broadens, and the maximum is shifted towards a higher time constant, eventually resulting in a bimodal distribution at larger d.

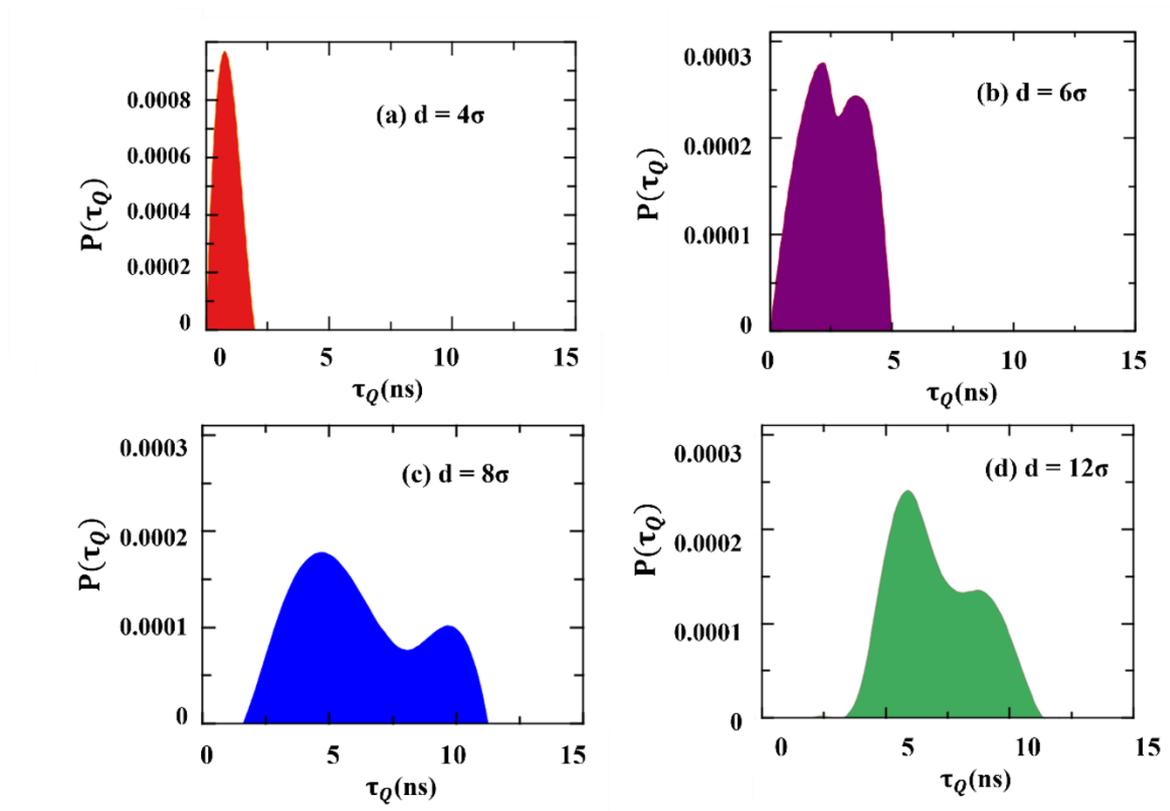

**Figure 10**. **Distribution of $\tau_Q$ at (a) d = 4σ, (b) d = 6σ, (c) = 8σ, and (d) d = 12σ. As the separation increases (from $d = 4$ σ to 12σ), the peak progressively broadens, and the maximum is shifted towards a higher time constant, eventually resulting in a bimodal distribution at larger d. Finally this leads to a bimodal distribution at larger values of d.**

Three primary factors compete significantly: (i) mutual diffusion of the polymer chain, (ii) mutual orientation of the polymers, and (iii) collapse. The bimodal distribution arises from two mechanisms; one of them is mutual translational diffusion of the center of mass, and the second one involves the rotational alignments of the two monomers.

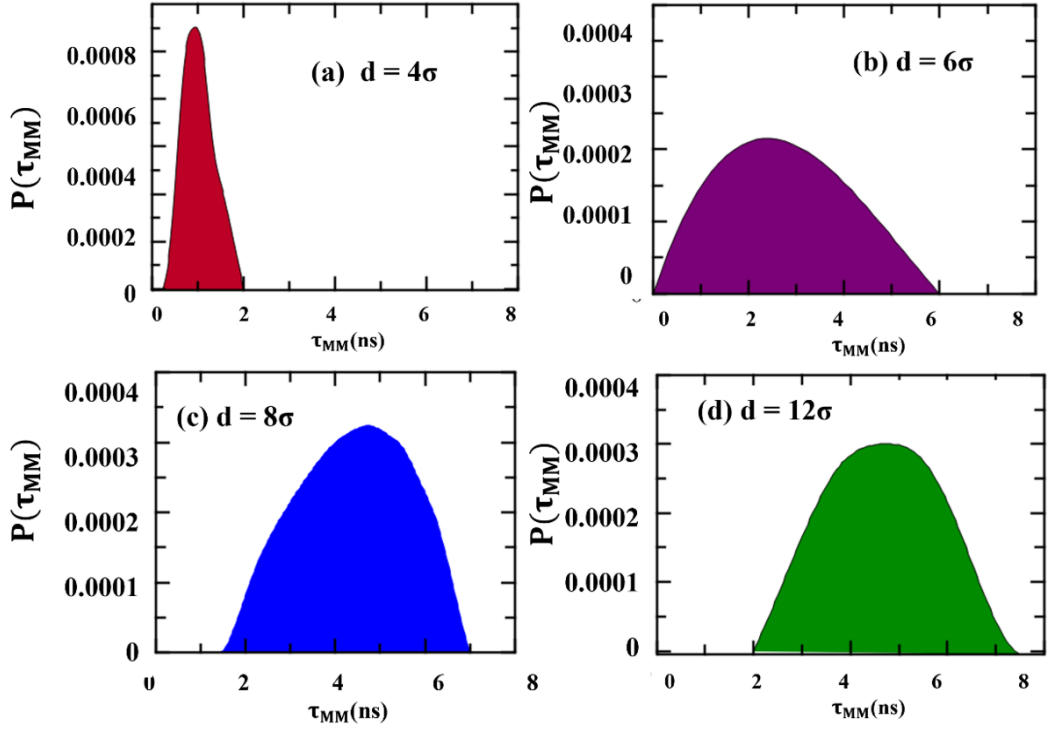

**Figure 11. Distribution of $\tau_{MM}$ at (a) d = 4σ, (b) d = 6σ, (c) = 8σ, and (d) d = 12σ. As the separation increases (from $d$ = 4 σ to 12σ), the peak progressively broadens, and the maximum is shifted towards a higher $\tau_{MM}$.**

The mechanism differs from small to large separation distances. At the limit of the large separation distance, the collapse of the monomer occurs first, followed by the mutual diffusion of the two collapsed globules. We find that studying the distribution of relaxation timescales is more revealing than comparing the average lifetime. Distribution of $\tau_{MM}$ is shown in **Figure 11**. It is observed that the peak progressively broadens with increasing the d, and the maximum is shifted towards a higher $\tau_{MM}$.

The reaction time distribution function can be written as an integration over the Heaviside function as follows.

$$P(\tau) = \int\int\int dR_{G_1} dR_{G_2} d\boldsymbol{R}_{MM} P(\boldsymbol{R}_{MM}, \tau) H(\boldsymbol{R}_{MM} - \delta)$$

(6)

$\mathbf{R_{MM}}$ is a three-dimensional position vector. $R_{G_1}$ and $R_{G_2}$ correspond to the radius of gyration of the two polymers and $\tau$ is the reaction time

H [ ] is the Heaviside function defined by,

$$H(R_{MM} - \delta) = 0, \ R_{MM} < \delta$$
$$= 1, \ R_{MM} > \delta \quad (7)$$

$\delta$ is the distance below which two polymers react, and the population of the monomeric state goes to 0. From our simulation, we observe that dimerization is essentially irreversible

## VIII. A Dynamical Disorder Model (DDM) of polymer dimerization

The trajectories calculated in the previous section depicted in various figures show the progression of different order parameters as functions of time under the influence of Langevin dynamics and interparticle interactions. Figure 11 depicts the reaction time distribution. This figure illustrates how the reaction time distribution evolves as the reaction progresses. The width of the distribution is a signature of the dynamical disorder present in the reaction system. It is a non-trivial task to formulate an analytical theory to describe such reaction time distribution. However, some progress can be made towards the formulation of such a theory, following the general approach suggested by Zwanzig in several papers.[50,51] Earlier, the dynamical disorder model was employed by Bagchi, Fleming, and Oxtoby[52] in the study of barrierless chemical reactions, Agmon and Hopfield[53] in the study of protein reactive dynamics, and Sumi and Marcus[54] in electron transfer reactions. The basic idea behind all these approaches is to introduce a reaction term in a generalized reaction-diffusion equation.

In this section, we employ the dynamical disorder model to capture the stochastic processes of collapse and dimerization. This theoretical framework provides invaluable insights into the complex dynamics and mechanisms underlying the process of polymer dimer formation. To gain a deeper understanding of the underlying mechanisms, the Dynamical Disorder Model provides a powerful theoretical framework. By considering the dynamic fluctuations and disorder within polymer chains, the Dynamical Disorder Model (DDM), while difficult to solve, offers a deeper understanding of the factors influencing dimerization kinetics and the structural properties of polymer dimers.

To obtain the time evolution of the probability distribution functions, we have closely followed Haken's elegant formulation,[55] also implemented by Zwanzig earlier [50]. Let us consider $C(R_{G1}, \mathbf{R_1}, t)$ is the time-dependent concentration of the monomeric polymer at position $\mathbf{R_1}$ with radius of gyration $R_{G1}$. The dimerization is a bimolecular reaction, involving two monomers. The decay of monomers happens through the said reaction. We assume an irreversible reaction. The rate equation of C(t), undergoing dimerization reaction, can be written as follows.

$$\frac{\partial C(R_{G1}, \mathbf{R_1}, t)}{\partial t} = -k(R_{G1}, R_{G2}, R_{MM}) C(R_{G1}, \mathbf{R_1}, t) C(R_{G2}, \mathbf{R_1} + \mathbf{R_{MM}}, t) \tag{8}$$

Where $R_{MM}$ is the separation between the center of the two monomers, $k(R_{G_1}, R_{G_2}, R_{MM})$ is the rate constant that depends on three factors, the two instantaneous radii of gyration $R_{Gi}$ of the respective monomers, and the distance of separation order parameter $R_{MM}(t)$.

The time-dependent formal solution of the rate equation Eq.(8) can be written as

$$C(R_{G1}, \mathbf{R_1}, t) = C(R_{G1}, \mathbf{R_1}, t=0) \int dR_{G_2} \int d\mathbf{R_2} \ \exp[-\int_0^t dt' \ k(t') C(R_{G_2}, \mathbf{R_2}, t')] \tag{9}$$

The complexity of the polymer dimerization problem, in comparison to a simpler small molecule association/dimerization reaction, arises from the conformational degrees of freedom of polymers. The situation is also different from binding of ligands to enzymes. In order to model the non-equilibrium conditions, we start with two vertical parallel polymer chains that are separated by a given distance $R_{MM}=d_0$.

Where we have made a change of coordinate by shifting the origin onto the centre of mass of the polymer 1. $R_2$ is now the distance of the centre of the monomeric polymer 2 from the centre of the monomer 1 which we earlier denoted as $R_{MM}$.

The probability distribution of the polymer in monomeric state, at time t is denoted by $P_1(R_{G1},t)$, where $R_G(t)$ is the time dependent radius the gyration, $R_{MM}(t)$ is the time dependent distance between monomeric polymer 1 and monomeric polymer 2. In this work we assume that the joint probability distributions can be decomposed as a product of two individual time dependent distribution functions, as below

$$C(R_{Gi}, R_{MM}, t) = g(R_{Gi},t) g(R_{MM},t) \qquad (10)$$

Eqs. (9) and (10) can be combined to obtain

$$C(R_{G_1},t) = C(R_{G_1}, t=0) \int dR_{G_2} \int d\mathbf{R}_{MM} \ \exp[-\int_0^t dt' k(R_{G_1}, R_{G_2}, \mathbf{R}_{MM}, t') g(R_{G_2},t') g(\mathbf{R}_{MM},t') \qquad (11)$$

The inter-monomer separation evolves with time, so does the rate of dimerization. Thus, the time dependent survival probability S(t) of the monomeric polymer can be given by

$$S(t) = < C(R_{G_1},t) > = \int dR_{G_1} C(R_{G_1},t) \qquad (12)$$

We model the time evolution of the probability distributions for $R_{Gi}(t)$ and $R_{MM}(t)$ as two independent diffusion processes in respective effective potentials. Additionally, we assume that both themselves as one-dimensional diffusion in the respective effective potential, and the equations of motion are given by the following two equations,

$$\frac{\partial g(R_{Gi},t)}{\partial t} = D_{G_i} \frac{\partial}{\partial R_{Gi}} \left( \frac{\partial}{\partial R_{Gi}} + V_{eff}(R_{Gi}) \right) g(R_{Gi},t) \qquad (13)$$

$$\frac{\partial g(R_{MM},t)}{\partial t} = D_M \frac{\partial}{\partial R_{MM}} \left( \frac{\partial}{\partial R_{MM}} + V_{MM}(R_{MM}) \right) g(R_{MM},t) \qquad (14)$$

Where, $D_{G_i}$ describes the conformational fluctuation of the i-th polymer as a diffusive process. The collapse occurs due to an effective potential $V_{eff}$, which can be regarded as a first approximation of a harmonic potential.

In the second equation, $D_M$ is the mutual diffusion constant that describes the separation between the two monomers. $V_{MM}$ is the effective potential between the two monomeric chains.

We can combine all of the above equations to obtain the following approximate expression.

$$S(t) = \iiint dR_{G_1} dR_{G_2} dR_{MM} \; e^{-k(R_{G_1},R_{G_2}R_{MM})t} g(R_{G_1},t) \; g(R_{G_2},t) g(R_{MM},t) \qquad (15)$$

We already mentioned that to avoid the complexity of the theory, we have approximated the two as separate one-dimensional processes. Furthermore, we have retained certain elements of the dynamic disorder model, but not solved the full equation of motion that would have required us to solve a complex three-variable joint probability distribution function. We believe that the approximate expression would be able to capture the essential elements of the nonequilibrium dimerization process.

In our nonequilibrium dimerization simulations, the initial conditions are well-defined

$$g(R_G,t=0) = \delta(R_G - a), \qquad (16)$$

$$g(R_{MM}, t=0) = \delta(R_{MM} - d_0) \tag{17}$$

Where a and d$_0$ are the radius of gyration of the polymer and center of mass distance between the two polymers at time t =0,

The probability distribution function of R$_{MM}$ is essentially the diffusion equation. [49]

$$g(R_{MM}, t) = \frac{1}{\sqrt{4\pi D_M t}} e^{-\frac{(R_{MM}(t) - R_{MM_0})^2}{4 D_M t}} \tag{18}$$

The probability distribution function of R$_G$ can be written as

$$g(R_{G_t}, t) = \frac{1}{\sqrt{2\pi \cdot (1 - \exp(-2bt)) \sigma_{R_G}^2}} \exp\left(-\frac{\left(R_{G_t} - \bar{R}_G\right)^2 - \left(R_{G_0} - \bar{R}_G\right)^2 e^{-bt}}{2(1 - \exp(-2bt)) \sigma_{R_G}^2}\right) \tag{19}$$

The time correlation between the values of R$_G$(t), observed at two times separated by t, can be written as

$$C_{R_G}(t) = \sigma_{R_G}^2 \exp(-bt) \tag{20}$$

Where $\sigma_{R_G}$ is the standard deviation, and b is the decay constant.

In the presence of dimerization reaction, where the final state is the dimer, the survival probability S(t) decay from 1 to 0. However, at equilibrium between the dimer and the monomer, the final value of S(t) is determined by the following equation.

$$k_{Eq} = \frac{C_{dimer}}{C_{mono}^2} \tag{21}$$

Where $C_{mono}$ and $C_{dimer}$ are the concentrations of polymer in a monomeric state and dimeric state, respectively.

## IX. Conclusions

The model, which has been widely used in theoretical studies, considers two straight vertical linear chains made up of Lennard Jones bead, placed initially (that is, at the time *t=0*) at a previously determined distance from each other in a parallel configuration. Our primary focus is to study the simultaneous occurrence of polymer dimerization and polymer collapse. Such a model can, of course, be generalized to more complex situations. In this study, we focussed on developing and partially implementing a theoretical description by introducing a number of order parameters that seem to play an important role in the polymer dimerization process. We observed that in this complex problem of polymer dimerization, there could be several candidates for order parameters. The requirement that the order parameters should be largely orthogonal to each other poses a problem. At large separations, the distance between the center of mass of the polymers and the radius of gyration serves the purpose well. At short separations, when the chains start to overlap in space, the contact order parameter needs to be considered. At large separations, the contact order parameter is zero. The number of contacts is found to grow sharply as the monomers come in contact. We have modified the standard definition of the contact order parameter by defining a contact in terms of the distance between atoms of the two monomeric polymers.

Numerical simulations reveal certain interesting features. They include (i) the role of the collapse of the individual monomeric proteins and the consequent decrease in the radius of gyration in these initially rod-like monomers, (ii) the seemingly existence of a critical distance which is much smaller than the length of monomeric proteins beyond which the monomers are likely to drift apart than associate, (iii) the large width of the reaction time distribution which points to the need of a dynamic disorder model in the spirit of the studies of Zwanzig and other people.

This work presents a dynamic disorder model, and the formulation can follow the established route where one combines the stochastic random walk or diffusion equation description with a parameter dependent reaction rate, a numerical solution is highly non-trivial because of the involvement of multiple order parameters in a partial differential equation.

Instead of attempting to solve these complex equations, we have followed the guidance of the dynamic disorder model to interrogate the progression of dimerization by calculating by simulations the time dependence of the order parameters along trajectories, and also calculate the reaction time distributions presented above. It is to be noted that, here we explore the limited parameter space where the equilibrium state of the monomeric single polymer chain is the collapsed state. In this study, we focus on polymers in poor solvents, where the final stable state is typically a collapsed form. However, in certain cases, such as with intrinsically disordered proteins (IDPs) that contain charged groups, water can act as a favorable solvent. Another example is β-amylin, where hydrophobic groups significantly contribute to polymer association. In these contexts, polymer association becomes crucial.[56–58]

Future work shall attempt to solve the dynamical disorder models. Also, a more realistic description of monomers is needed. Extension to the association of real protein is a problem that has not been studied in great detail. Previous studies conducted in this laboratory investigated the dissociation of insulin dimers, uncovering several noteworthy features. [13,59,60] Extending that study to association is highly desirable but also highly non-trivial.

## Acknowledgement

We thank Professor Parbati Biswas, Dr. Sarmistha Sarkar, Dr. Shubham Kumar, and Dr. Subhajit Acharya for useful discussions. B.B. thanks the Science and Engineering Research Board (SERB), India, for the National Science Chair Professorship and the Department of Science and Technology, India, for partial research funding. S.M. thanks IISc. for a research fellowship.

## Data Availability

The data that support the findings of this study are available within the article.